\documentclass[twocolumn,showpacs,preprintnumbers,amsmath,amssymb,latexsym,prl,footinbib,floatfix,superscriptaddress]{revtex4-2}

\usepackage{amsmath,amssymb,latexsym,slashed,multirow}

\usepackage{graphicx}
\newcommand{\f}[2]{\frac{#1}{#2}}
\newcommand{\tf}[2]{{\textstyle \frac{#1}{#2}}}

\newcommand{\de}{\partial}
\newcommand{\la}{\langle}
\newcommand{\ra}{\rangle}

\newcommand{\Oc}{{\cal O}}
\renewcommand{\Re}{{\rm Re}}

\newcommand{\Tt}{{\cal T}}
\newcommand{\tr}{{\rm tr}\,}
\newcommand{\csqrt}[1]{\sqrt[\mathbb{C}]{#1}}
\newcommand{\sgnepsilon}{\varepsilon}
\newcommand{\Dmeas}{\mathcal{D}}
\newcommand{\erf}{{\rm erf}}

\begin{document}

\title{Lattice simulations of the QCD chiral transition at real baryon density}

\author{Szabolcs Bors\'anyi} \affiliation{Department of Physics,
  Wuppertal University, Gaussstr.\ 20, D-42119, Wuppertal, Germany}

\author{Zolt\'an Fodor} 
  \affiliation{Department of Physics, Wuppertal University, Gaussstr.\ 20, D-42119, Wuppertal, Germany}
  \affiliation{Pennsylvania State University, Department of Physics, State College, Pennsylvania 16801, USA}
  \affiliation{J{\"u}lich Supercomputing Centre, Forschungszentrum J{\"u}lich, D-52425 J{\"u}lich, Germany}
  \affiliation{ELTE E\"otv\"os Lor\'and University, Institute for
  Theoretical Physics, P\'azm\'any P\'eter s\'et\'any 1/A, H-1117, Budapest,
  Hungary}

\author{Matteo Giordano}
\affiliation{ELTE E\"otv\"os Lor\'and University, Institute for
  Theoretical Physics, P\'azm\'any P\'eter s\'et\'any 1/A, H-1117, Budapest,
  Hungary}

\author{S\'andor D.\ Katz}
\affiliation{ELTE E\"otv\"os Lor\'and University, Institute for
  Theoretical Physics, P\'azm\'any P\'eter s\'et\'any 1/A, H-1117, Budapest,
  Hungary}
\affiliation{
  MTA-ELTE Theoretical Physics Research Group,
  P\'azm\'any P\'eter s\'et\'any 1/A, H-1117 Budapest, Hungary.}

\author{D\'aniel N\'ogr\'adi}
\affiliation{ELTE E\"otv\"os Lor\'and University, Institute for
  Theoretical Physics, P\'azm\'any P\'eter s\'et\'any 1/A, H-1117, Budapest,
  Hungary}

\author{Attila P\'asztor}
\email{Corresponding author: apasztor@bodri.elte.hu}
\affiliation{ELTE E\"otv\"os Lor\'and University, Institute for
  Theoretical Physics, P\'azm\'any P\'eter s\'et\'any 1/A, H-1117, Budapest,
  Hungary}

\author{Chik Him Wong} \affiliation{Department of Physics, Wuppertal
  University, Gaussstr.\ 20, D-42119, Wuppertal, Germany}

\begin{abstract}
  State-of-the-art lattice QCD studies of hot and dense strongly
  interacting matter currently rely on extrapolation from zero or
  imaginary chemical potentials. The ill-posedness of numerical
  analytic continuation puts severe limitations on the reliability of
  such methods.
  Here
  we use the more direct sign reweighting method to perform lattice
  QCD simulation of the QCD chiral transition at finite real baryon
  density on
  phenomenologically relevant lattices.
  This method does not require analytic continuation and avoids the
  overlap problem associated with generic reweighting schemes, 
  so has
  only statistical but no uncontrolled systematic
  uncertainties for a fixed lattice setup. 
  This opens up a new window to study hot and dense strongly interacting matter from first principles. 
  We perform simulations up to a baryochemical potential-temperature
  ratio of $\mu_B/T=2.5$ covering most of the RHIC Beam Energy Scan
  range in the chemical potential. We also clarify the connection of
  the approach to the more traditional phase reweighting method.
\end{abstract}

\maketitle

\paragraph{Introduction}

The properties of strongly interacting matter at high temperature and
density play a role in a variety of issues, such as the early history
of the Universe and the scattering of heavy ions.  These issues are
currently at the center of intense theoretical and experimental
investigations, and a deeper understanding of hot and dense strongly
interacting matter would greatly help in furthering progress.  In
particular, the chiral transition has garnered a lot of
interest~\cite{Aoki:2006we, Borsanyi:2010bp, HotQCD:2019xnw,
  Kotov:2021rah}, as the comparison of theoretical predictions with
results from heavy-ion experiments can potentially challenge our
understanding of strong interactions based on Quantum Chromodynamics
(QCD). It is therefore important to obtain predictions for the
behavior of strongly interacting matter near the chiral transition
starting from first principles.

The most well established method for first-principles studies of QCD
in the strongly coupled regime near the transition is lattice
QCD~\cite{Montvay:1994cy}.  The lattice approach turns the path
integral of quantum field theory into a practical numerical method by
mapping it to a statistical-mechanics system.  This method can in
principle be systematically improved to reach arbitrary accuracy.  QCD
at finite baryon density is, however, not amenable to first-principle
lattice studies using standard techniques, since in this case the
Boltzmann weights in the path integral representation are complex and
so not suitable for importance-sampling algorithms.  A variety of
methods have been proposed over the years to side-step this complex
action problem. None of these methods is, however, completely
satisfactory, as they all suffer from systematic effects of some kind.
Methods based on using an imaginary chemical
potential~\cite{deForcrand:2002hgr,DElia:2002tig,DElia:2009pdy,
  Cea:2014xva,Bonati:2014kpa,Cea:2015cya,Bonati:2015bha,
  Bellwied:2015rza,DElia:2016jqh,Gunther:2016vcp,Alba:2017mqu,
  Vovchenko:2017xad,Bonati:2018nut,Borsanyi:2018grb,Bellwied:2019pxh,
  Borsanyi:2020fev} or a Taylor expansion around vanishing chemical
potential~\cite{Allton:2002zi,Gavai:2003mf,Gavai:2004sd,Allton:2005gk,
  Gavai:2008zr,MILC:2008reg,Borsanyi:2011sw,Borsanyi:2012cr,
  Bellwied:2015lba,Ding:2015fca,Bazavov:2017dus,HotQCD:2018pds,
  Giordano:2019slo,Bazavov:2020bjn} involve a certain amount of
modeling, as they necessarily make assumptions about the functional
dependence of physical observables on the chemical potential, in order
to reconstruct them at real, finite chemical potential.  Despite its
formal exactness, the overlap problem when reweighting from zero
chemical potential $\mu_B=0$~\cite{Hasenfratz:1991ax,Barbour:1997ej,
  Fodor:2001au,Fodor:2001pe,Fodor:2004nz,Giordano:2019gev} makes it
very difficult to quantify statistical and systematic uncertainties. 
This is also true for the complex Langevin
approach~\cite{Seiler:2012wz,Sexty:2013ica,Aarts:2014bwa,
  Fodor:2015doa,Sexty:2019vqx,Kogut:2019qmi,Scherzer:2020kiu} due to
its convergence issues.  Yet other speculative methods, such
as dual variables~\cite{Gattringer:2014nxa,Marchis:2017oqi} or
Lefshetz thimbles~\cite{Cristoforetti:2012su,Cristoforetti:2013wha,
  Alexandru:2015xva,Alexandru:2016lsn,Nishimura:2017vav} have only
been successfully used to study toy models so far.

Although technically manifesting as different, the analytic
continuation problem of the Taylor and imaginary chemical potential
methods and the overlap problem of reweighting from $\mu_B=0$ have the
same origin: an inability to directly sample the gauge configurations
most relevant to finite-density QCD, thus requiring an extrapolation
that hopefully captures the features of the theory of interest.  One
would instead like to perform simulations in a theory from which
reconstruction of the desired theory is the least affected by
systematic effects, by (\textit{i}) keeping as close as possible to
the most relevant configurations, thus minimizing the overlap problem,
and by (\textit{ii}) making the complex-action problem, or sign
problem, due to cancellations among contributions, as mild as
possible.  A method satisfying both requirements -- ``sign
reweighting'' -- has sporadically been mentioned in the literature for
quite some time~\cite{deForcrand:2002pa,deForcrand:2009zkb,
  Hsu:2010zza,Alexandru:2005ix,Li:2010qf,Li:2011ee,Giordano:2020roi}.
In fact, as reweighting is reduced to a sign factor only, the overlap
problem is absent. 
Moreover, sign reweighting is the optimal choice, with the weakest
sign problem, out of reweighting schemes based on simulating theories
where the Boltzmann weights differ from the desired ones only by a
function of the phase of the quark
determinant~\cite{deForcrand:2002pa,deForcrand:2009zkb,Hsu:2010zza}.
This approach is so far the closest one can get to sampling the most
relevant configurations according to the original, sign-problem-ridden
path integral, and allows one to answer detailed questions about the
gauge configurations that determine the nature of dense strongly
interacting matter.

While optimal, the ``sign quenched'' theory that one has to simulate
in the sign reweighting approach is unfortunately not a local field
theory, so that the standard algorithms of lattice QCD do not apply.
This leads to more costly numerics, and has prevented so far the use
of sign reweighting in large scale simulations on fine lattices.  The
state-of-the-art so far was the study on toy lattices of
Ref.~\cite{Giordano:2020roi}.  After further optimization, here we
demonstrate that sign reweighting has become viable for
phenomenologically relevant lattices.  We perform simulations of the
sign quenched theory with 2-stout improved staggered fermions at
$N_\tau=6$ -- a lattice action that is often used (at zero or
imaginary chemical potential) as the first point of the continuum
extrapolation for thermodynamic
quantities~\cite{Aoki:2006we,Aoki:2006br,Borsanyi:2010cj,Bali:2011qj,
  Bali:2012zg,Borsanyi:2015yka,Bonati:2015bha,Brandt:2017oyy,
  Bonati:2018nut,DElia:2019iis}.  We therefore obtain results directly
(up to reweighting by a sign) at a finite real chemical potential, up
to a baryochemical potential-temperature ratio of
$\hat{\mu}_B=\frac{\mu_B}{T}=2.5$, which is near the upper end of the
chemical potential range of the RHIC Beam Energy
Scan~\cite{STAR:2017sal,Bzdak:2019pkr,STAR:2020tga} and is already in
a region of the phase diagram where analytic continuation methods stop
being predictive.  Beyond previous results on toy lattices, this is the first result 
in the literature obtained at real baryon density without
any of the unknown systematic uncertainties, such as those coming from
the overlap problem and analytic continuation.
To aid further studies of this kind we also provide a way to
estimate the severity of the sign problem -- the main bottleneck for
sign reweighting studies -- based on susceptibility measurements at $\mu_B=0$.

\paragraph{The overlap problem and sign reweighting}

A generic reweighting method reconstructs expectation values in a
desired target theory, with microscopic variables $U$, path integral
weights $w_t(U)$, and partition function
$Z_t = \int\Dmeas U \ w_t(U)$, using simulations in a theory with real
and positive path integral weights $w_s(U)$ and partition function
$Z_s = \int\Dmeas U \ w_s(U)$, via the formula:
\begin{equation}
  \label{eq:reweight}
  \left\langle \Oc \right\rangle_t = \frac{\left\langle \frac{w_t}{w_s}
      \Oc \right\rangle_s }{\left\langle
      \frac{w_t}{w_s}\right\rangle_s}\,,\quad
  \langle \Oc \rangle_x = \f{1}{Z_x} \int\Dmeas U \ w_x(U) \Oc(U)\,,
\end{equation}
where $x$ may stand for $t$ or $s$. 
When the target theory is lattice QCD at finite chemical potential,
the target weights $w_t(U)$ have wildly fluctuating phases: this is
the infamous sign problem of lattice QCD. In addition to this problem,
generic reweighting methods also suffer from an overlap problem: the
probability distribution of the reweighting factor $w_t/w_s$ has
generally a long tail, which cannot be sampled efficiently 
in standard Monte Carlo simulations.
It is actually the overlap problem, rather than the sign problem, that
constitutes the immediate bottleneck in QCD when one tries to extend
reweighting results
to finer lattices~\cite{Giordano:2020uvk}. 

\begin{figure*}[t]
  \centering
  \includegraphics[width=0.45\textwidth]{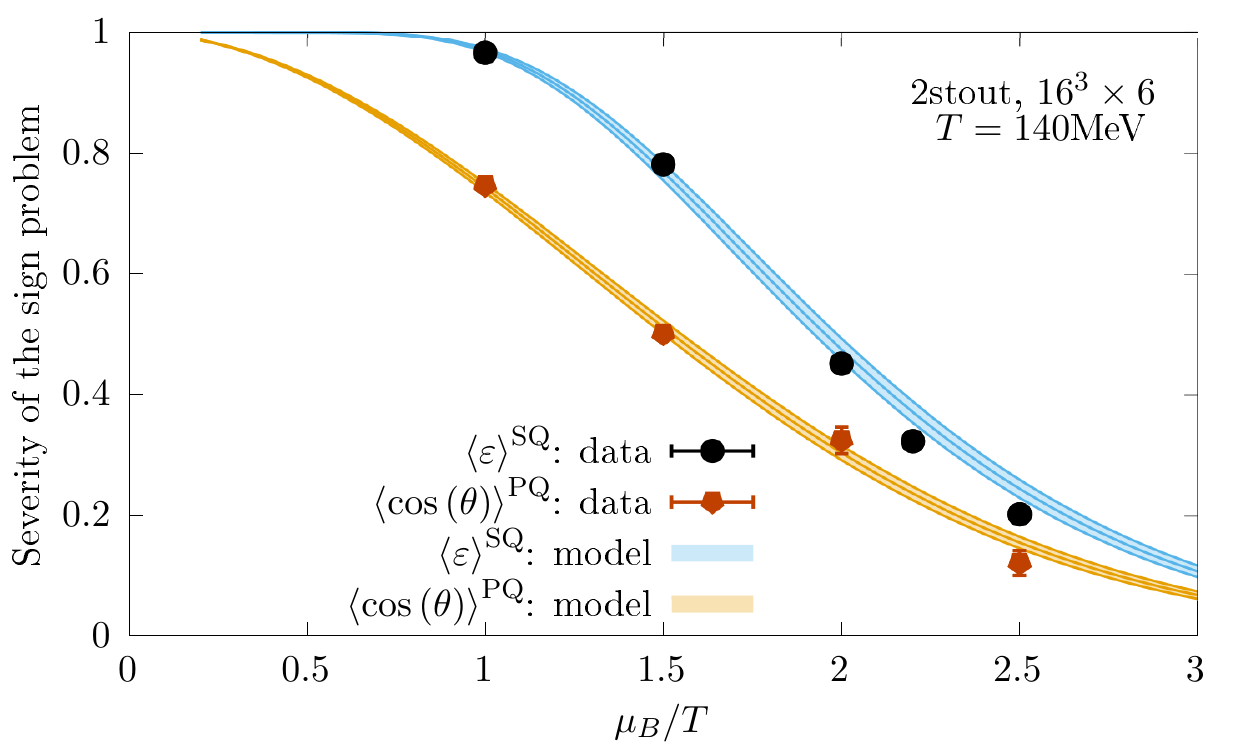}
  \includegraphics[width=0.45\textwidth]{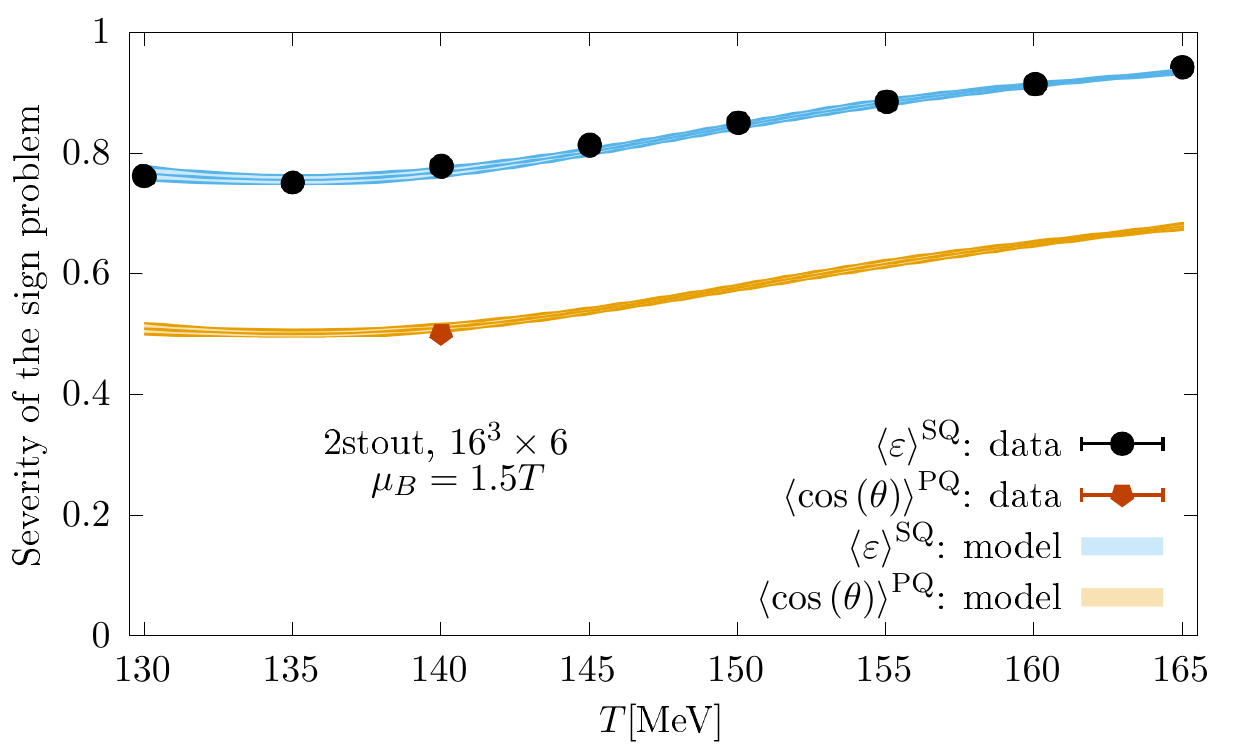}
  \caption{The strength of the sign problem as a function of $\mu_B/T$
    at $T=140$ MeV (left) and as a function of $T$ at $\mu_B/T=1.5$. A
    value close to $1$ shows a mild 
    sign problem. A small value
    indicates a severe sign problem.  Data for sign reweighting
    (black) and phase reweighting (orange) are from direct
    simulations.  Predictions of the Gaussian model are also shown.}
  \label{fig:sign}
\end{figure*}

A way to address the overlap problem is to utilize sign
reweighting~\cite{deForcrand:2002pa,
  deForcrand:2009zkb,Hsu:2010zza,Alexandru:2005ix,Li:2010qf,Li:2011ee,Giordano:2020roi}.
The partition function of lattice QCD is real due to charge
conjugation invariance, and at finite temperature $T$ and finite real
quark chemical potential $\mu$ one can write 
\begin{equation}
  \label{eq:Redet}
  \begin{aligned}
    Z(T,\mu) &= \int \Dmeas U\,\Re\det M(U,\mu)  e^{-S_g(U)}\,,
    \end{aligned}
\end{equation}
where $S_g$ is the gauge action, $\det M$ denotes the fermionic
determinant, including all quark types with their respective mass
terms, as well as rooting in the case of staggered fermions, and the
integral is over all link variables $U$. Replacing the determinant
with its real part is not permitted for arbitrary expectation values,
but it is allowed for observables satisfying $\Oc(U^*) = \Oc(U)$, as
well as for those obtained as derivatives of $Z$ with respect to real
parameters, such as the chemical potential or the quark mass.  As most
important observables in bulk thermodynamics are of this kind, one can
use Eq.~\eqref{eq:Redet} as the starting point for a reweighting
scheme.  Denoting by $\sgnepsilon$ the sign of $\Re\det M(U,\mu)$, one
has
\begin{align}
    & Z(T,\mu)  = \la \sgnepsilon\ra^{\rm SQ}_{T,\mu} Z_{\rm SQ}(T,\mu)
    \,, \nonumber\\
    &    Z_{\rm SQ}(T,\mu)= \int \Dmeas U\,|\Re\det M(U,\mu)|
      e^{-S_g(U)}\,,   \label{eq:sr4}\\
    &    \la \Oc\ra^{\rm SQ}_{T,\mu} = \! 
    \f{1}{Z_{\rm SQ}(T,\mu)}\!\int\! \Dmeas U\,\Oc(U)
    |\Re\det M(U,\mu)|
    e^{-S_g(U)}\,.    \nonumber
\end{align}
Here SQ stands for ``sign quenched'' and $Z_{\rm SQ}$ defines the
``sign-quenched ensemble''.  The desired expectation values are then
obtained by setting $w_s=|\Re\det M(U,\mu)| e^{-S_g(U)}$,
$w_t=\Re\det M(U,\mu) e^{-S_g(U)}$ and $w_t/w_s=\sgnepsilon$ in
Eq.~\eqref{eq:reweight}.  Since $\sgnepsilon = \pm 1$, reweighting
boils down to a sign factor, and one avoids the problem of inaccurate
sampling of the tails of the probability distribution of the
reweighting factor (i.e., the overlap problem), since the tails are
absent by construction.  The only 
problem left is the sign problem, which is under control as long as
$\la \sgnepsilon\ra^{\rm SQ}_{T,\mu}$ is safely not zero within
errors. In this case, sign reweighting gives reliable results, and
unlike any other of the commonly used methods for $\mu_B$, error bars
(for a fixed lattice setup) are statistical only.

\paragraph{Severity of the sign problem}

A key step in addressing the feasibility of our approach is estimating
the severity of the sign problem. The sign reweighting approach is
closely related to the better known phase reweighting
approach~\cite{Fodor:2007vv,Endrodi:2018zda}, where in
Eq.~\eqref{eq:reweight} we have $w_t=\det M(U,\mu)e^{-S_g(U)}$ and
$w_s=|\!\det M(U,\mu)| e^{-S_g(U)}$, which defines the phase quenched
ensemble ${\rm PQ}$.  In this ensemble the severity of the sign
problem is measured by the average phase factor
$ \la e^{i\theta} \ra^{\rm PQ}_{T,\mu}= \la \cos \theta \ra^{\rm
  PQ}_{T,\mu}$, while in the SQ ensemble it is measured by
$\la \sgnepsilon\ra^{\rm SQ}_{T,\mu}= {\la\cos \theta\ra^{\rm
    PQ}}/{\la |\!\cos \theta| \ra^{\rm PQ}}$.  Clearly,
$\la \cos \theta \ra^{\rm PQ}_{T,\mu} \le \la \sgnepsilon \ra^{\rm
  SQ}_{T,\mu}$, 
so the sign problem is generally weaker in the SQ case.  Moreover, the
probability distribution of the phases $\theta= \arg \det M$ in the
phase quenched theory, $P_{\rm PQ}(\theta)$, controls the strength of
the sign problem in both ensembles.  A simple quantitative estimate
can then be obtained with the following two-step approximation:
(\textit{i}) in a leading order cumulant expansion,
$P_{\rm PQ}(\theta)$ is assumed to be a wrapped Gaussian distribution;
(\textit{ii}) the chemical potential dependence of its width is
approximated by the leading order Taylor expansion,
$\sigma(\mu)^2 \approx \left\langle \theta^2 \right\rangle_{\rm LO}
=-\frac{4}{9} \chi^{ud}_{11} \left( LT \right)^3\hat{\mu}_B^2
$~\cite{Allton:2002zi}, where
$\chi^{ud}_{11} = \f{1}{T^2}\frac{\partial^2 p }{\partial
  \mu_u \partial \mu_d
  }|_{\mu_u=\mu_d=0}$
is the disconnected part of the light
quark susceptibility, obtained in $\mu=0$ simulations.  In this
approximation both cases can be calculated analytically, with
$\la \cos \theta \ra^{\rm PQ}_{T,\mu} \approx
e^{-\frac{\sigma^2(\mu)}{2}}$ in the phase quenched case, while in the
sign quenched case the expression for
$\la \sgnepsilon\ra^{\rm SQ}_{T,\mu}$ is more involved.  It is worth
noting the different asymptotics of the two cases.  The small-$\mu$
(i.e., small-$\sigma$) asymptotics are notably very different, with
$ \la \cos\theta\ra^{\rm PQ}_{T,\mu} \sim 1 - \f{\sigma^2(\mu)}{2}$ analytic
in $\hat{\mu}_B$,
while in the sign quenched case
$\la \sgnepsilon\ra^{\rm SQ}_{T,\mu}$ is not analytic,
\begin{equation}
  \la \sgnepsilon\ra^{\rm SQ}_{T,\mu}
    \underset{\hat{\mu}_B\to  0}{\sim}
  1 -\left(\tf{4}{\pi} \right)^{\f{5}{2}}
  \left(\tf{\sigma^2(\mu)}{2}\right)^{\f{3}{2}} e^{-\f{\pi^2}{8
      \sigma^2(\mu)}}\,,
\end{equation}
approaching 1 faster than any polynomial (see the supplemental
material for a derivation).  The large-$\mu$ or large volume
asymptotics are on the other hand quite similar: in the large-$\sigma$
limit a wrapped Gaussian tends to the uniform distribution, and so at
large chemical potential or volume one arrives at
\begin{equation}
  \label{eq:asyratio}
  \f{\la \sgnepsilon\ra^{\rm SQ}_{T,\mu}}{\la \cos\theta\ra^{\rm PQ}_{T,\mu}}
  \underset{\hat{\mu}_B\,\text{or}\, V\to \infty}{\sim}
  \left(
    \int_{-\pi}^{\pi}d\theta\, |\!\cos \theta| \right)^{-1}=\frac{\pi}{2} \,, 
\end{equation}
which asymptotically translates to a factor of
$(\f{\pi}{2})^2 \approx 2.5$ less statistics needed for a sign
quenched as compared to a phase quenched simulation.  

We compare our Gaussian model with simulation results for both the
sign reweighting and phase reweighting approach in
Fig.~\ref{fig:sign}.  Error bars on the model come solely from the
statistical errors of $\chi^{ud}_{11}$ at $\mu_B=0$.  Our model
describes reasonably well our simulation data at small $\mu$ in both
cases, deviating less than $1\sigma$ from the actual measured strength
of the sign problem up to $\hat{\mu}_B=2$.  While deviations are
visible at larger $\mu$, even at the upper end of our $\hat{\mu}_B$
range the deviation is at most $25\%$, and Eq.~\eqref{eq:asyratio}
approximates well the relative severity of the sign problem in the two
ensembles at $\hat{\mu}_B > 1.5$.

In summary, this shows that we can estimate the severity of the sign
problem using $\mu_B=0$ simulations only, making the planning of such
reweighting studies practical.  Furthermore we have also demonstrated
- using simulations at real chemical potential - that at an aspect
ratio of $LT \approx 2.7$ the sign problem is manageable up to
$\hat{\mu}_B = 2.5$.
Covering the range of the RHIC Beam Energy Scan is
therefore feasible.

\begin{figure*}[t]
  \centering
  \includegraphics[width=0.45\textwidth]{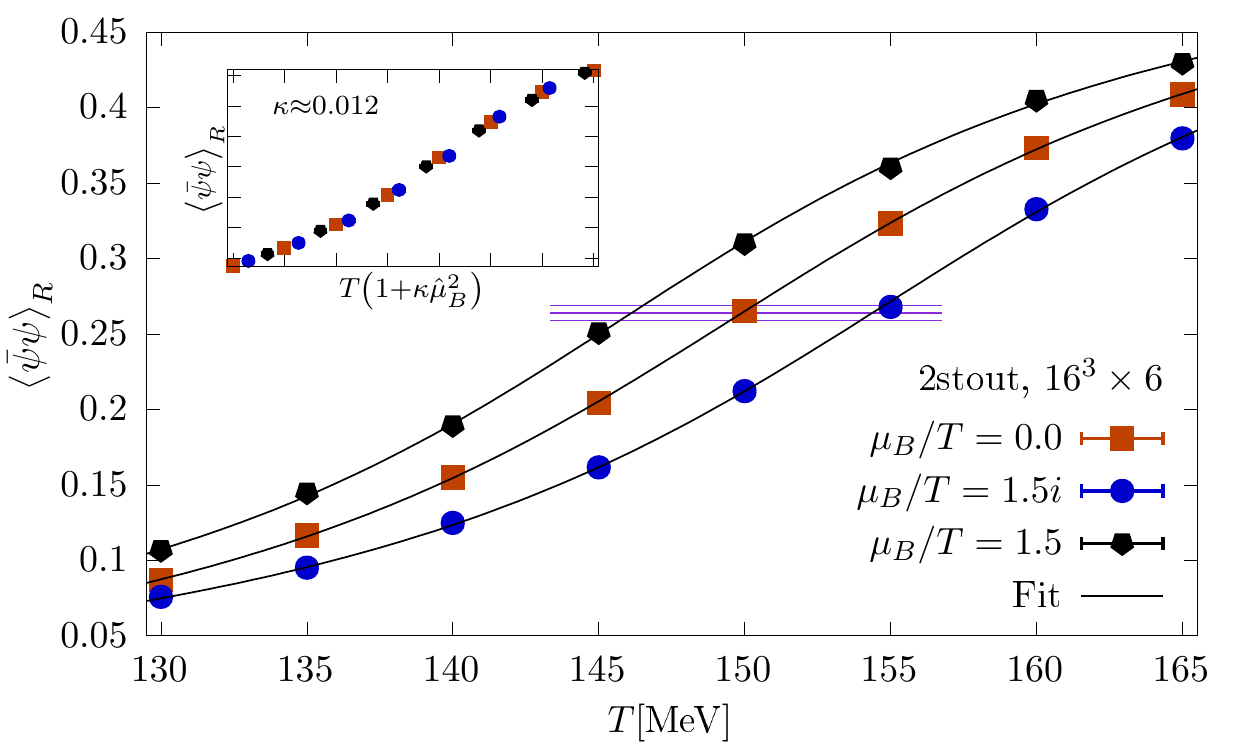}
  \includegraphics[width=0.45\textwidth]{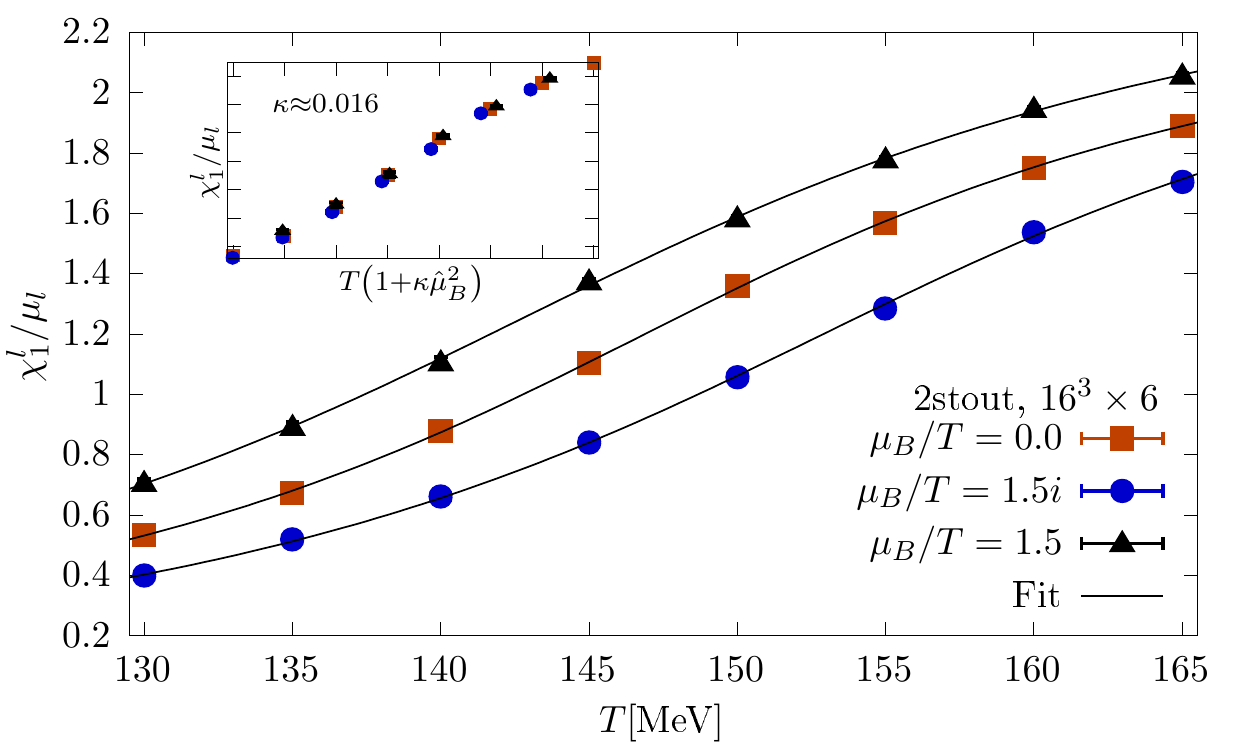}
  \caption{The renormalized chiral condensate (left) and the light
    quark number-to-light quark chemical potential ratio (right) as a
    function of temperature at $\mu_B/T=1.5$. 
    The datapoints are shown together with an arcotangent based fit.
    In the insets, collapse
    plots are shown in the variable
    $T \cdot (1+\kappa \left(\f{\mu_B}{T}\right)^2)$, with
    $\kappa \approx 0.012$ for the chiral condensate and
    $\kappa \approx 0.016$ for the quark number. In the left panel the
    value of the condensate at the crossover temperature at $\mu_B=0$
    is also shown.}
  \label{fig:Tscan}
\end{figure*}

\paragraph{Simulation setup}

We simulated the sign quenched ensemble using 2+1 flavors of rooted
staggered fermions. We used a tree-level Symanzik improved gauge
action, and two steps of stout smearing~\cite{Morningstar:2003gk} with
$\rho=0.15$ on the gauge links fed into the fermion determinant, with
physical quark masses, using the kaon decay constant $f_K$ for scale
setting (see Ref.~\cite{Aoki:2009sc} for details).  We studied
$16^3\times 6$ lattices at various temperatures $T$ and light-quark
chemical potential $\mu_u=\mu_d=\mu_l=\mu=\mu_B/3$ with a zero strange quark
chemical potential $\mu_s=0$, corresponding to a strangeness chemical
potential $\mu_S=\mu_B/3$.  We performed a scan in chemical potential
at fixed $T=140\,{\rm MeV}$, and a scan in temperature at fixed
$\hat{\mu}_B= 1.5$.  Simulations were performed by modifying the RHMC
algorithm at $\mu_B=0$ by including an extra accept/reject step that
takes into account the factor $\f{|\Re \det M(\mu)|}{\det M(0)}$.  The
determinant was calculated with the reduced matrix
formalism~\cite{Hasenfratz:1991ax} and dense linear algebra, with no
stochastic estimators involved.  See the supplemental material for
more details.

\paragraph{Observables}

We now proceed to display physics results.  The light-quark chiral
condensate was obtained via the formula
\begin{equation}
  \label{eq:chiralcondensate}
  \begin{aligned}
    \la\bar{\psi}\psi\ra_{T,\mu} &= \f{1}{Z(T,\mu)}\f{\de
      Z(T,\mu)}{\de
      m_{ud}} \\
    &=\f{T}{V}\f{1}{\la\sgnepsilon\ra^{\rm SQ}_{T,\mu}}\left\la
      \sgnepsilon\f{\de}{\de m_{\rm ud}} \ln\left |\Re \det M\right|
    \right\ra^{\rm SQ}_{T,\mu} \,,
  \end{aligned}
\end{equation}
with the determinant $\det M=\det M(U,m_{ud},m_s,\mu)$ calculated in
the reduced matrix formalism at different light-quark masses and fed
into a symmetric difference,
$\f{df(m)}{dm} \approx \f{f(m+\Delta m)-f(m-\Delta m)}{2\Delta m}$,
choosing $\Delta m$ small enough to make the systematic error from the
finite difference negligible compared to the statistical error.  The
renormalized condensate was obtained with the prescription
\begin{equation}
  \label{eq:chiralcondensate3}
  \la \bar{\psi}\psi\ra_R(T,\mu) =   -\f{m_{ud}}{f_\pi^4}\left[
    \la \bar{\psi}\psi\ra_{T,\mu} -\la \bar{\psi}\psi\ra_{0,0}
  \right]\,.
\end{equation}
We also calculated the light quark density 
\begin{equation}
\begin{aligned}
  \chi^l_1 &\equiv \f{\partial \left(p/T^4\right)}{\partial \left(
      \mu/T \right) } = \frac{1}{VT^3} \f{1}{Z(T,\mu)} \f{\partial
    Z(T,\mu)}{\partial \hat{\mu}} \\ 
  &= \f{1}{VT^3 \la\sgnepsilon\ra^{\rm SQ}_{T,\mu}} \left\la
    \sgnepsilon\f{\de}{\de \hat{\mu}} \ln\left |\Re \det M\right|
  \right\ra^{\rm SQ}_{T,\mu} \,,
\end{aligned}
\end{equation}
evaluating the derivative analytically using the reduced matrix
formalism (see the supplemental material).

\begin{figure*}[t]
  \centering
  \includegraphics[width=0.45\textwidth]{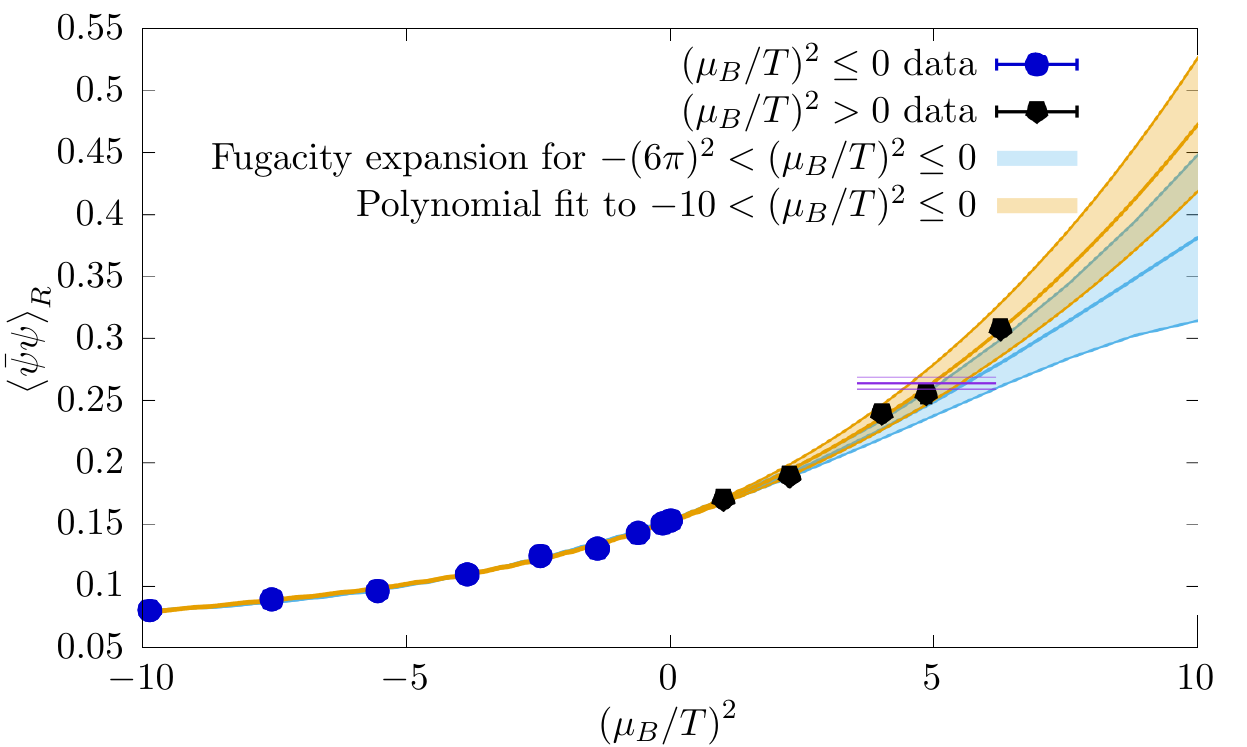}
  \includegraphics[width=0.45\textwidth]{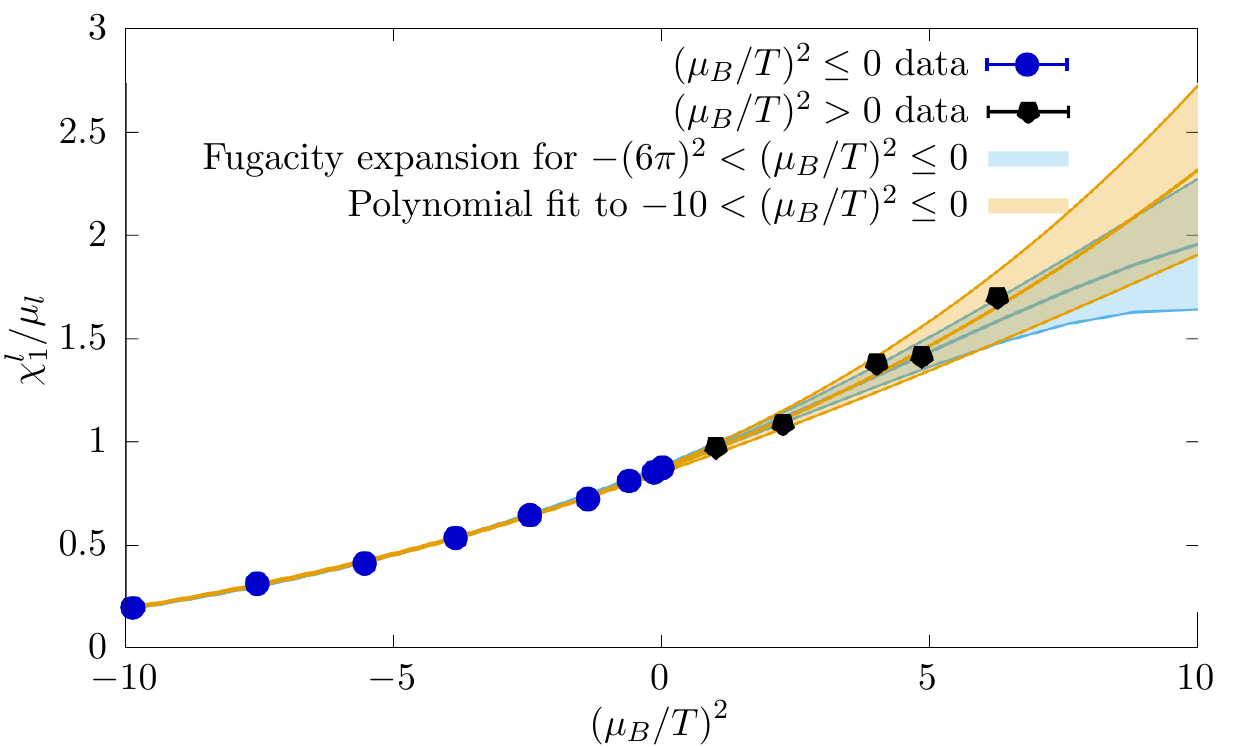}
  \caption{The renormalized chiral condensate (left) and the light
    quark number-to-light quark chemical potential ratio (right) as a
    function of $\left( \mu_B/T \right)^2$ at temperature $T=140$
    MeV. Data from simulations at real $\mu_B$ (black) are compared
    with analytic continuation from imaginary $\mu_B$ (blue).  In the
    left panel the value of the condensate at the crossover
    temperature at $\mu_B=0$ is also shown.  The simulation data cross
    this line at $\mu_B/T \approx 2.2$.}
  \label{fig:muscan}
\end{figure*}

\paragraph{Temperature scan}

Our results for a temperature scan between $130$ MeV and $165$ MeV at
real chemical potential $\hat{\mu}_B=1.5$, zero chemical potential,
and imaginary chemical potential $\hat{\mu}_B=1.5i$ are shown in
Fig.~\ref{fig:Tscan}. The most important quantitative question 
one can
address with such a temperature scan is the strength of the crossover
transition. 
Methods based on analytic continuation cannot address this
particular issue efficiently.
It was in fact demonstrated by numerical
simulations that for imaginary chemical potentials the strength of the
transition is to a good approximation constant~\cite{Borsanyi:2020fev,
  Borsanyi:2021sxv}. However, the extrapolation of such a behavior to
real chemical potentials is inherently dangerous. It is usually
assumed that the transition at physical masses and $\mu_B=0$ is close
to the O(4) scaling regime in the continuum 
theory~\cite{Pisarski:1983ms,Butti:2003nu,Pelissetto:2013hqa,Grahl:2013pba,Nakayama:2014sba} (or O(2) with
staggered fermions on the lattice~\cite{Boyd:1991fb}), while close to the critical
endpoint one expects to see $\mathbb{Z}_2$ scaling.  One then cannot
assess at what point one enters the $\mathbb{Z}_2$ region using gauge
configurations that are only sensitive to O(4) (or O(2)) criticality,
and extrapolations from such configurations are very likely to miss a
transition to the other regime -- even if it exists.  Our results,
however, show that the transition is not getting any stronger up to
$\hat{\mu}_B=1.5$, as convincingly demonstrated by the collapse
plot in the inset of Fig.~\ref{fig:Tscan}. In fact, data at
$\hat{\mu}_B = 0,1.5,1.5i$ are all reasonably well described by one 
and the same function of $T(1+\kappa \hat{\mu}_B^2)$.

\paragraph{Chemical potential scan}

Our results for the chemical potential scan at a fixed temperature of
$T=140$ MeV are shown in Fig.~\ref{fig:muscan}.  We have performed
simulations at 
$\hat{\mu}_B=1, 1.5, 2, 2.2, 2.5$.
The point at $\hat{\mu}_B=2.2$
corresponds roughly to the chiral transition, as at this
point the chiral condensate is close to its value at the $\mu_B=0$
crossover. 

The sign-quenched results are compared with the analytic continuation
from imaginary chemical potential results, obtained by extrapolating
suitable fits to the imaginary-$\mu_B$ data from negative to positive
$\hat{\mu}_B^2$.
We considered two types of fits.  (\textit{i}) As the simplest
ansatz, we fitted the data with a cubic polynomial in
$\hat{\mu}_B^2$
in the range
$\hat{\mu}_B^2 \in [-10,0]$.
(\textit{ii}) As an alternative, we
also used suitable ans\"atze for
$\left\langle \bar{\psi}\psi\right\rangle_R$ condensate and
$\chi^l_1/\hat{\mu}_l$ based on the fugacity expansion
$p/T^4 = \sum_n A_n \cosh(n \hat{\mu})$, fitting the data in the
entire imaginary-potential range
$\hat{\mu}_B^2 \in \left[ -(6 \pi)^2,0\right]$
using respectively 7 and
6 fitting parameters.  Fit results are also shown in
Fig.~\ref{fig:muscan}; only statistical errors are displayed.  While
sign reweighting and analytic continuation 
give compatible results, at the upper half of the
$\mu_B$ range the errors from sign reweighting are an order of
magniture smaller.  In fact, sign reweighting can penetrate the region
$\hat{\mu}_B>2$ where the extrapolation of many quantities is not yet
possible~\cite{Bazavov:2017dus,Borsanyi:2020fev}.

\paragraph{Summary and outlook}

We have demonstrated that sign reweighting has become a viable
approach to finite-density lattice QCD.  This is the first lattice
study performed with a phenomenologically relevant lattice action
(2-stout improved staggered fermions, 6 time slices, aspect ratio
$LT \approx 2.7$) that does not require analytic continuation, unlike
the Taylor expansion and imaginary $\mu_B$ methods, and is free from
the overlap problem of more traditional reweighting approaches.  We
also presented a way to estimate the severity of the sign problem from
$\mu_B=0$ simulations, making the method practical: the computational
cost for a given $\mu_B$ and a given lattice action is now easily
predictable.

Our temperature scan at $\mu_B/T=1.5$ shows no sign of the transition
getting stronger. Furthermore, while the results of the $\mu_B$ scan
at $T=140$ MeV are compatible with those obtained from extrapolation
from imaginary $\mu_B$, the errors of the sign reweighting method are
an order of magnitude smaller, opening up new possibilities.

Our chemical potential scan shows that small statistical errors can be
achieved up to $\mu_B/T=2.5$, and our temperature scan shows that the
severity of the sign problem is only weakly dependent on the
temperature (Fig.~\ref{fig:sign}, right).  Our method is then
optimized enough to make a full scan of the chiral transition region
in the RHIC Beam Energy Scan range feasible, with computing resources available
today.  Such a scan allows us to attack the most important open question of the Beam Energy
Scan, and decide whether the crossover transition becomes stronger in
the range, as expected for a nearby critical
endpoint~\cite{Bzdak:2019pkr,Shuryak:2019ikv,STAR:2020tga,
  Shuryak:2020yrs,Braun-Munzinger:2020jbk,Mroczek:2020rpm}. It would
also allow us to obtain the equation of state directly, and test the
range of validity of several recently proposed resummation
schemes~\cite{Borsanyi:2021sxv,Mondal:2021jxk} for the Taylor
expansion of the pressure in $\mu_B$.

The lattice action used in this study is often 
the first point
of a continuum extrapolation in QCD thermodynamics.  Furthermore,
while the sign problem is exponential in the physical volume, it is
not so in the lattice spacing. Continuum-extrapolated finite $\mu_B$
results in the range of the RHIC Beam Energy Scan are then almost
within reach for the phenomenologically relevant aspect ratio of
$LT \approx 3$.  On the theoretical side, sampling the most relevant
configurations allows one to study detailed aspects of the theory at
$\mu_B>0$, such as spectral statistics of the Dirac operator, likely
leading to new insights.

\paragraph{Acknowledgements}
We thank Tam\'as G. Kov\'acs for useful discussions.
The project was supported by the BMBF Grant No. 05P18PXFCA.
This work was also supported by the Hungarian National Research,
Development and Innovation Office, NKFIH grant KKP126769.
A.P. is supported by the J. Bolyai Research
Scholarship of the Hungarian Academy of Sciences and by the \'UNKP-20-5 New
National Excellence Program of the Ministry for Innovation and Technology.
The authors gratefully acknowledge the Gauss Centre for Supercomputing e.V.
(www.gauss-centre.eu) for funding this project by providing computing time on the
GCS Supercomputers JUWELS/Booster and JURECA/Booster at FZ-Juelich.

\newpage

\section*{Supplemental Material}
\label{sec:SM}

\subsection*{A. Analytic estimates of the strength of the sign
  problem}
\label{sec:severity}

In this section we discuss the strength of the sign problem, both in
the sign quenched and in the phase quenched reweighting methods.

\subsubsection*{A.1. Distribution of the determinant phase}
\label{sec:detphase}

The partition functions $Z$ of the target theory, i.e., QCD,
$Z_{\rm PQ}$ of the phase quenched ensemble, and $Z_{\rm SQ}$ of the
sign quenched ensemble read
\begin{equation}
  \label{eq:partfunc}
  \begin{aligned}
    Z & = \int \Dmeas U\,\det M e^{-S_g} = \int \Dmeas U\,\Re\det M
    e^{-S_g}  \,,\\
    Z_{\rm PQ} & = \int \Dmeas U\,|\det M| e^{-S_g}  \,,\\
    Z_{\rm SQ} & = \int \Dmeas U\,|\Re\det M| e^{-S_g}  \,,
  \end{aligned}
\end{equation}
where $\det M$ denotes the fermionic
determinant, including all quark types with their respective mass
terms, as well as rooting in the case of staggered fermions.
The strength of the sign problem is measured by the ratio of partition
functions appearing in the reweighting procedure. These are
conveniently expressed in terms of the distribution
$P_{\rm PQ}(\theta)$ of the phase of the determinant
$\theta=\arg\det M$ in the phase quenched theory,
\begin{equation}
  \label{eq:partfunc2}
  \begin{aligned}
    &   P_{\rm PQ}(\theta) =    \la \delta\left(\arg(\det M) -
      \theta\right)\ra^{\rm PQ} 
    \\ &= \f{1}{Z_{\rm PQ}}  \int \Dmeas U\,|\!\det M|
    e^{-S_g[U]}
    \delta(\arg\det M - \theta)\,.
  \end{aligned}
\end{equation}
One finds
\begin{equation}
  \label{eq:partfunc3}
  \begin{aligned}
    \f{Z}{Z_{\rm PQ}} &= \f{1}{Z_{\rm PQ}} \int \Dmeas U\,\cos\theta
    |\!\det M|  e^{-S_g[U]} \\
    &= \int_{-\pi}^{\pi} d \theta\, P_{\rm PQ}(\theta) \cos \theta  = \la
    \cos\theta \ra^{\rm PQ} \,,
  \end{aligned}
\end{equation}
for the phase quenched approach, and
\begin{equation}
  \label{eq:partfunc4}
  \begin{aligned}
    \f{Z}{Z_{\rm SQ}} &= \f{Z}{Z_{\rm PQ}} \left(\f{1}{Z_{\rm PQ}}
      \int \Dmeas U\, |\!\cos\theta| |\!\det M| e^{-S_g[U]}\right)^{-1}
    \\
    &= \frac{
      \int_{-\pi}^{\pi} d \theta\, P_{\rm PQ}(\theta) \cos \theta 
    }{
      \int_{-\pi}^{\pi} d \theta\, P_{\rm PQ}(\theta) |\!\cos \theta| }
    = \f{\la \cos\theta \ra^{\rm PQ}}{ \la |\!\cos\theta| \ra^{\rm
        PQ}}=\la \sgnepsilon\ra^{\rm SQ}\,,
  \end{aligned}
\end{equation}
for the sign quenched approach.
Since $ \la |\!\cos\theta| \ra^{\rm PQ}\le 1$, the sign problem in the
sign-quenched theory is generally less severe than
in the phase-quenched theory.

\subsubsection*{A.2. Polar decomposition of the determinant}
\label{sec:poldec}

The fermion determinant can be written as
$\det M(U,\mu) = e^{i\Phi(U,\mu) + V(U,\mu)}$ with real functions
$\Phi, V$. 
Due to the properties $M(U,\mu)^\dag = M(U,-\mu^*)$ and
$M(U,\mu)^* =M(U^*,\mu^*)$, one has for $\mu\in\mathbb{R}$
\begin{equation}
  \label{eq:detpolardec}
  \begin{aligned}
    e^{-i\Phi(U,\mu) + V(U,\mu)}&= e^{i\Phi(U,-\mu) + V(U,-\mu)}\\ &=
    e^{i\Phi(U^*,\mu) + V(U^*,\mu)}\,,
  \end{aligned}
\end{equation}
implying
\begin{equation}
  \label{eq:detpolardec2}
  \begin{aligned}
    \Phi(U,-\mu) &= -\Phi(U,\mu) =         \Phi(U^*,\mu)\,, \\
    V(U,-\mu)&= V(U,\mu) = V(U^*,\mu)\,.
  \end{aligned}
\end{equation}
In summary, $\Phi$ is $C$-odd and $\mu$-odd, so at least of order
$O(\mu)$, while $V$ is $C$-even and $\mu$-even.

\subsubsection*{A.3. Gaussian approximation}
\label{sec:gapp}

In a cumulant expansion, the complex-phase average in the
phase-quenched theory at finite $\mu$ reads in the lowest-order
(Gaussian) approximation,
\begin{equation}
  \label{eq:PQphase}
  \la\cos\theta\ra^{\rm PQ} = \la e^{i\Phi}\ra^{\rm PQ} =
  e^{-\f{1}{2}\la\Phi^2\ra^{\rm PQ} + \ldots} = e^{O(\mu^2)}\,.
\end{equation}
In this approximation the phase in the phase-quenched theory obeys a
wrapped normal distribution~\cite{fisher_1993} centered at zero,
\begin{equation}
  \label{eq:sprob_wg}
  \begin{aligned}
  P_{\rm PQ}(\theta)
  &\underset{\substack{\text{Gaussian}\\\text{approx.}}}{=}
  \f{1}{\sqrt{2\pi}\sigma}\sum_{n=-\infty}^\infty 
  e^{-\f{1}{2\sigma^2}(\theta + 2\pi n )^2} \\ &= \f{1}{2\pi}\sum_{n=-\infty}^\infty
  e^{-n^2\f{\sigma^2}{2} + i n\theta} \\ &=\f{1}{2\pi}\sum_{n=-\infty}^\infty
  e^{-n^2\f{\sigma^2}{2}}\cos n\theta \,.
  \end{aligned}
\end{equation}
Expanding around $\mu=0$, we now find
\begin{equation}
  \label{eq:PQfrommuzero2}
  \begin{aligned}
    \la\Phi(U,\mu)^2\ra^{\rm PQ}&=\f{\left\la
        \Phi(U,\mu)^2e^{V(U,\mu)-V(U,0)}\right\ra_0}{\left\la
        e^{V(U,\mu)-V(U,0)}\right\ra_0}\\
    &=\mu^2\left\la \Phi'(U,0)^2\right\ra_0 + O(\mu^4)\,,
  \end{aligned}
\end{equation}
where $\la\ldots\ra_0$ is the expectation value at $\mu=0$, and
moreover~\cite{Allton:2002zi} 
\begin{equation}
  \label{eq:PQfrommuzero3}
  \begin{aligned}
    & \left\la \Phi'(U,0)^2\right\ra_0 = \left\la \left[\tr\left(
          M(0)^{-1} M'(0)\right)\right]^2\right\ra_0 \\ &= N_f^2
    \f{\de^2\log Z(\mu_u,\mu_d)}{\de\mu_u\de\mu_d}
    \bigg|_{\mu_u=\mu_d}\\
    &= N_f^2 VT \f{\de^2\f{p}{T^4}}{\de\f{\mu_u}{T}\de\f{\mu_d}{T}}
    \bigg|_{\mu_u=\mu_d} = 4VT\chi_{11}^{ud}\,,
  \end{aligned}
\end{equation}
where $N_f$ denotes the number of degenerate quark flavors coupled to the same chemical 
potential $\mu$. In the Gaussian approximation and to lowest order in $\mu$, one has then
\begin{equation}
  \label{eq:PQfrommuzero4}
  \la\cos\theta\ra^{\rm PQ}
  \underset{\substack{\text{Gaussian}\\ \text{approx.}}}{=} e^{-\f{1}{2}\sigma^2(\mu)}
  \underset{\text{LO}}{=}   e^{-\f{1}{2}
    VT^3 \f{4}{9}\chi_{11}^{ud}\f{\mu_B^2}{T^2}}\,, 
\end{equation}
where $\sigma^2=\left\langle \Phi^2 \right\rangle^{\mathrm{PQ}}$ and $\mu_B = 3\mu$ is the baryochemical potential.

\subsubsection*{A.4. Sign problem in the sign-quenched theory}
\label{sec:sgnprob}

We can now estimate the severity of the sign problem in the
sign-quenched theory in the Gaussian approximation.  Using the first
line in Eq.~\eqref{eq:sprob_wg} one finds the exact expression
\begin{equation}
  \label{eq:sprob_wg2}
  \begin{aligned}
{\cal N}(\sigma) &=    \int_{-\pi}^{\pi} d\theta\, P_{\rm
  PQ}(\theta)\left|\cos\theta\right| \\ 
    &=\f{1}{\sqrt{2\pi}\sigma}\sum_{n=-\infty}^\infty (-1)^n
    \int_{n\pi-\f{\pi}{2}}^{n\pi+\f{\pi}{2}}
        d\theta\,\cos\theta \,  e^{-\f{1}{2\sigma^2}\theta^2} \\
        &=\f{e^{- \f{\sigma^2}{2}}}{2}\sum_{n=-\infty}^\infty (-1)^n \left(
          f_n(\sigma)-f_{n-1}(\sigma) \right) \\
        &= e^{- \f{\sigma^2}{2}} \big[ f_0(\sigma) + (f_0(\sigma)-f_1(\sigma))  \\
        &\quad\quad\quad\quad\quad -(f_1(\sigma)-f_2(\sigma))+ \dots \big]\,, 
  \end{aligned}
\end{equation}
where 
\begin{equation}
  \label{eq:sprob_wg3}
  \begin{aligned}
    f_n(\sigma) &= \Re\,
    \erf\left(\tf{\pi}{\sqrt{2}\sigma}\left(n+\tf{1}{2}\right)+i\tf{\sigma}{\sqrt{2}}\right)\,,\\
    \erf(z) &= \f{2}{\sqrt{\pi}}\int_0^z dt\,e^{-t^2}\,,
  \end{aligned}
\end{equation}
and we used the property $f_{-n} = - f_{n-1}$.
From this result one can obtain the behavior at small $\sigma$ using
the asymptotic expansion of the error function $\erf(z)$ 
(see, e.g.,
Ref.~\cite{abramowitz1964handbook}). To lowest order,
\begin{equation}
  \label{eq:sprob_wg_asy}
  \begin{aligned}
    {\cal N}(\sigma)e^{\f{1}{2}\sigma^2}
    &\underset{\sigma\to
      0}{\sim}
    1 +
    \f{4}{\pi}\left(\f{2\sigma^2}{\pi}\right)^{\f{3}{2}}
    e^{-\f{\pi}{4}\f{\pi}{2\sigma^2}}\,,
    \\
    \la\sgnepsilon\ra^{\rm SQ} 
    \underset{\substack{\text{Gaussian}\\ \text{approx.}}}{=} 
    \f{e^{-\f{1}{2}\sigma^2}}{{\cal
        N}(\sigma)} &
    \underset{\sigma\to
      0}{\sim} 1 -
    \f{4}{\pi}\left(\f{2\sigma^2}{\pi}\right)^{\f{3}{2}}
    e^{-\f{\pi}{4}\f{\pi}{2\sigma^2}}\,,
  \end{aligned}
\end{equation}
with neglected contributions of order
$O(\sigma^5e^{-{\pi^2}/{8\sigma^2}})$ and
$O(\sigma^3e^{- {\pi^2}/{\sigma^2}})$.  To study the asymptotic
large-$\sigma$ behavior it is more convenient to use the third line in
Eq.~\eqref{eq:sprob_wg} to get
\begin{equation}
  \label{eq:sprob_wg4}
  \begin{aligned}
    {\cal N}(\sigma) &= \f{1}{2\pi}\sum_{n=-\infty}^\infty
    e^{-n^2\f{\sigma^2}{2}} \int_0^{2\pi} d\theta\,|\!\cos\theta|\cos
    n\theta \\ &= \f{2}{\pi}\sum_{n=-\infty}^\infty
    e^{-4n^2\f{\sigma^2}{2}} \int_0^{\f{\pi}{2}}
    d\theta\,\cos\theta\cos 2n\theta \\ 
    &= -\f{2}{\pi}\sum_{n=-\infty}^\infty \f{ (-1)^n
      e^{-4n^2\f{\sigma^2}{2}}}{4n^2-1} \,.
  \end{aligned}
\end{equation}
At large $\sigma$,
${\cal N}(\sigma) \simeq \f{2}{\pi}(1+\tf{2}{3}e^{-2\sigma^2})$, and
so
\begin{equation}
  \label{eq:larges}
  \begin{aligned}
    {\cal N}(\sigma) &\simeq
    \f{2}{\pi}(1+\tf{2}{3}e^{-2\sigma^2})\,,\\
    \f{\la\sgnepsilon\ra^{\rm SQ} }{ \la\cos\theta\ra^{\rm PQ}} &=
    \f{1}{{\cal N}(\sigma)}\underset{\sigma\to\infty}{\to} \f{\pi}{2}
    \,.
  \end{aligned}
\end{equation}
Note that the asymptotic ratio $\f{\pi}{2}$ is not specific to
the Gaussian model, and depends only on $P_{\rm PQ}(\theta)$
approaching a uniform distribution in the large $\mu$ or large volume
limit. The correction term of order $e^{-2\sigma^2}$ is instead
specific to the Gaussian model.

\subsection*{B. Algorithmic details}
\label{sec:algo}

In this paper we have employed the sign-reweighting method to study
finite-density QCD using rooted staggered fermions. Here we discuss
the details of the formulation and of the simulation algorithm.

\subsubsection*{B.1. Staggered rooting at finite chemical potential}
\label{sec:stag}

For QCD with two degenerate light quarks $u$ and $d$, and a heavier
strange quark $s$, coupled respectively to chemical potentials
$\mu_u = \mu_d = \mu$ and $\mu_s=0$, one has for the partition
function with rooted staggered fermions
\begin{equation}
  \label{eq:sr_rs_1}
  \begin{aligned}
    Z(T,\mu) &= \int \Dmeas U\, [\det M_{\rm
      stag}(U,m_{ud},\mu)]^{\f{1}{2}} \\ &\phantom{= \int_T \Dmeas U\,
    }\times[\det M_{\rm stag}(U,m_s,0)]^{\f{1}{4}} e^{-S_g[U]} \\ &=
    \int \Dmeas U\, \Re\{ [\det M_{\rm
      stag}(U,m_{ud},\mu)]^{\f{1}{2}}\}\\ &\phantom{= \int_T \Dmeas
      U\, }\times[\det M_{\rm stag}(U,m_s,0)]^{\f{1}{4}} e^{-S_g[U]}
    \,.
  \end{aligned}
\end{equation}
Here $ M_{\rm stag}(U,m,\mu) = m + D_{\rm stag}(U,\mu)$ with
\begin{equation}
  \label{eq:sr_rs_2}
  D_{\rm stag}(U,\mu) =\f{1}{2}\sum_{\alpha=1}^4\eta_\alpha
  ( e^{\mu\delta_{\alpha 4}} U_\alpha \Tt_\alpha -
  e^{-\mu\delta_{\alpha 4}} \Tt_\alpha^\dag U_\alpha^\dag )
\end{equation}
where $\eta_\alpha$, $U_\alpha$ and $\Tt_\alpha$ denote respectively
the staggered phases, the link variables, and the translation operator
in direction $\alpha$. Our choice of quark chemical potentials
corresponds to baryochemical potential $\mu_B=3\mu$ and strangeness
chemical potential $\mu_S= \mu_B/3$. The integral in
Eq.~\eqref{eq:sr_rs_1} is over all SU(3) link variables in a
$N_s^3\times N_\tau$ hypercubic lattice with the SU(3) Haar measure.
Periodic boundary conditions in all directions are assumed for the
link variables.  Antiperiodic boundary conditions in the temporal
direction and periodic boundary conditions in the spatial directions
for fermions are implicitly included in the definition of
$\Tt_\alpha$.

The partition function for the sign-quenched ensemble and the
corresponding expectation values are
\begin{equation}
  \label{eq:sr_rs_5}
  \begin{aligned}
    Z_{\rm SQ}(T,\mu) &= \int DU\, {\cal R}(U,\mu) e^{-S_g[U]}\,,\\
    \la \Oc\ra^{\rm SQ}_{\mu} &= \f{1}{Z_{\rm SQ}(\mu)} \int
    DU\,\Oc(U) {\cal R}(U,\mu) e^{-S_g[U]}\,,
  \end{aligned}  
\end{equation}
where 
\begin{equation}
  \label{eq:sr_rs_5bis}
  \begin{aligned}
    {\cal R}(U,\mu)&= |\Re\{ [\det M_{\rm
      stag}(U,m_{ud},\mu)]^{\f{1}{2}}\}|\\
    & \phantom{= \f{1}{Z_{\rm SQ}(\mu)}\int }\times [\det M_{\rm
      stag}(U,m_s,0)]^{\f{1}{4}}\,.
  \end{aligned}
\end{equation}
The square root of the determinant is generally ambiguous at finite
$\mu$, and the simple behavior under charge conjugation used to get
the second equality in Eq.~\eqref{eq:sr_rs_1} is not guaranteed to
hold. In this paper, as in Refs.~\cite{Fodor:2001pe,Fodor:2004nz,
  Giordano:2020roi}, we adopt the following prescription.  The fermion
determinant at finite $\mu$ can be written as follows in terms of the
$\mu$-independent eigenvalues $\Lambda_i$ of the reduced
matrix~\cite{Hasenfratz:1991ax},
\begin{equation}
  \label{eq:redmat1}
  \det M_{\rm stag}(U,m,\mu) =
  e^{3V\hat{\mu}}\prod_{i=1}^{6V}\left[\Lambda_i(U,m) -
    e^{-\hat{\mu}}\right]\,, 
\end{equation}
where $\hat{\mu}=\f{\mu}{T}$. Clearly,
\begin{equation}
  \label{eq:redmat2}
  \begin{aligned}
    &\det M_{\rm stag}(U,m,\mu) \\ &= \det M_{\rm stag}(U,m,0) \f{
      \det M_{\rm stag}(U,m,\mu)}{
      \det M_{\rm stag}(U,m,0)} \\
    &= \det M_{\rm stag}(U,m,0)\prod_{i=1}^{6V}
    \left(\f{\Lambda_i(U,m) e^{\f{\hat{\mu}}{2}} -
        e^{-\f{\hat{\mu}}{2}}}{\Lambda_i(U,m) - 1 }\right) \,,
  \end{aligned}
\end{equation}
with $\det M_{\rm stag}(U,m,0)$ real and positive. We now set
\begin{equation}
  \label{eq:redmat3}
  \begin{aligned}
    & \protect{[\det M_{\rm stag}(U,m,\mu)]^{\f{1}{2}}} \\ &\equiv
    \sqrt{\det M_{\rm stag}(U,m,0)} \prod_{i=1}^{6V}
    \csqrt{\f{\Lambda_i(U,m) e^{\f{\hat{\mu}}{2}} -
        e^{-\f{\hat{\mu}}{2}}}{\Lambda_i(U,m) - 1 }} \,,
  \end{aligned}
\end{equation}
where the complex square root $\csqrt{z}$ appearing on the right-hand
side is defined as the analytic continuation of the positive
determination of the real square root with a branch cut on the
negative real axis. This choice and Eq.~\eqref{eq:redmat3} fully
specify the rooting procedure.  Notice that by construction
$[\det M_{\rm stag}(U,m,0)]^{\f{1}{2}}=\sqrt{\det M_{\rm
    stag}(U,m,0)}=\csqrt{\det M_{\rm stag}(U,m,0)}$. Since with our
choice $\csqrt{z^*}=\csqrt{z}^*$, and since
the sets of eigenvalue of complex conjugate gauge configurations
satisfy
$\{\Lambda_i(U^*,m)\} =\{\Lambda_i(U,m)^*\}$, the rooted determinant
obeys
\begin{equation}
  \label{eq:redmat4}
  [\det M_{\rm
    stag}(U^*,m,\mu)]^{\f{1}{2}}=    \left([\det M_{\rm
    stag}(U,m,\mu)]^{\f{1}{2}}\right)^* \,,
\end{equation}
so that reality of $Z$ follows from charge-conjugation invariance, and
the second equality in Eq.~\eqref{eq:sr_rs_1} holds.

\subsubsection{B.2. Simulation algorithm}
\label{sec:simu}

Simulating the sign-quenched ensemble is a nontrivial task, since even
assuming that a pseudofermion representation exists, it does not seem
easy to find. As in Ref.~\cite{Giordano:2020roi}, we have then split
this task in two parts.  The fermionic part ${\cal R}(U,\mu)$ of the
Boltzmann weight, Eq.~\eqref{eq:sr_rs_5bis}, can be identically recast
as
${\cal R}(U,\mu) = \f{{\cal R}(U,\mu)}{{\cal R}(U,0)}{\cal R}(U,0)$,
where $ {\cal R}(U,0)$ is the usual rooted staggered determinant at
$\mu=0$, while by construction $\f{{\cal R}(U,\mu)}{{\cal R}(U,0)}$
reads
\begin{equation}
  \label{eq:sr_rs_8}
  \f{{\cal R}(U,\mu)}{{\cal R}(U,0)} =\left|\Re \prod_{i=1}^{6V}
    \csqrt{\f{\Lambda_i(U,m_{ud}) e^{-\f{\hat{\mu}}{2}} -
        e^{\f{\hat{\mu}}{2}}}{\Lambda_i(U,m_{ud}) - 1 }}\right|\,.
\end{equation}
The factor ${\cal R}(U,0)$ can be simulated using a standard RHMC
algorithm; including the $\mu$-dependent ratio in the accept/reject
step at the end of the RHMC trajectories, one obtains the desired
Boltzmann distribution for the sign-quenched ensemble.  Notice that
since only the absolute value of the real part of the determinant is
involved, in Eq.~\eqref{eq:sr_rs_8} one can ignore the sign ambiguity
inherent in the rooting procedure, and so
$\f{{\cal R}(U,\mu)}{{\cal R}(U,0)}$ can be evaluated more simply and
more efficiently by separately computing the full fermionic
determinants at zero and finite $\mu$ and taking any of their square
roots, instead of computing $\{\Lambda_i\}$. Calculation of the
eigenvalues is needed only when measuring observables and the
reweighting factor.

When simulating the phase quenched ensemble, we pursue a similar
approach, with the sign quenched factor
$\f{|\Re \det M^{1/2}(\mu)|}{\det M(0)}$ being substituted by
$\f{| \det M^{1/2}(\mu)|}{\det M(0)}$. This way we can avoid the
introduction of an explicit symmetry breaking parameter $\lambda$ -
coupled to the charged pion field - as is usual for phase quenched
simulations~\cite{Kogut:2002zg}. This gets rid of the need to do a
$\lambda \to 0$ extrapolation, and the high numerical cost associated
with 
diagonalization of the reduced matrix at each $\lambda$.

The most computationally 
expensive part of the simulations is the
diagonalization of the reduced matrix, 
whose cost is dominated
by reduction of the matrix to upper Hessenberg form, which
asymptotically takes $\f{10}{3} (6N_s^3)^3$ floating point
operations~\cite{GoluVanl96}. The scaling of the determinant
calculations is the same up to a prefactor, as Gaussian elimination
takes asymptotically $\f{1}{6}(6 N_s^3)^3$ floating point
operations~\cite{GoluVanl96}. The theoretical ratio of the two costs
is therefore $20$. This, however, does not seem to be true in
practice, due to more optimizations available for Gaussian
elimination.  On a modern GPU, with the publicly available MAGMA
linear algebra library~\cite{tdb10}, one diagonalization for our
$16^3 \times 6$ lattices is approximately $50$ times
more expensive than
one Gaussian elimination. With the measurements taking place after every
$16$ Monte Carlo updates, in order to sufficiently decorrelate the configurations, the cost of 
measurements was roughly $60\%$ of the entire computational cost.

\subsection{C. Statistics tables}
\label{sec:stat}

The statistics of our numerical simulations are summarized in
Tab.~\ref{tab:stat}.  For the real chemical potential runs, each
configuration is separated by 16 Monte Carlo updates; for the zero and
imaginary chemical potential runs, each configuration is separated by
10 RHMC trajectories.

\begin{table}[th]
    \begin{center}
    \begin{tabular}{|c|c|}
\hline
    \multicolumn{2}{|c|}{$\mu_B/T$ scan, $T=140$MeV}                               \\
\hline
\hline
$\mu_B/T$                                           & $N_{\rm configs}/1000$  \\  
\hline
0                                                  & 7.3                     \\  
\rule{0pt}{4pt} ${\rm i}\,\frac {6 \pi}{46}\cdot k$                        & \multirow{2}{*}{3.5}                     \\  
\rule{0pt}{4pt}     $(k=1,2,\dots,46)$                             &                         \\  
1.0                                                & 12                      \\  
1.5                                                & 11                      \\  
2.0                                                & 15                      \\  
2.2                                                & 12                      \\  
2.5                                                   & 12                        \\  
\hline
\end{tabular}
\hfil    \begin{tabular}{|c|c|}
\hline
 \multicolumn{2}{|c|}{$T$ scan, $\mu_B/T=1.5$} \\
\hline
\hline
$T[{\rm MeV}]$   & $N_{\rm configs}/1000$    \\  
\hline
      \rule{0pt}{8pt}     130              & 9                         \\  
           135              & 11                        \\  
           140              & 11                        \\  
           145              & 11                        \\  
           150              & 11                        \\  
           155              & 12                        \\  
           160              & 10                        \\  
           165              & 12                        \\  
\hline
\end{tabular}

\end{center}
\caption{Summary of our statistics for the chemical potential scan
  (left) and the temperature scan (right).}
    \label{tab:stat}
\end{table}

\providecommand{\noopsort}[1]{}\providecommand{\singleletter}[1]{#1}%
%

\end{document}